\documentclass[times, twoside, watermark]{zHenriquesLab-StyleBioRxiv}
\usepackage{blindtext}
\usepackage{lipsum}
\usepackage{graphicx}

\leadauthor{Percannella} 

\begin{document}

\title{A multi-task neural network for atypical mitosis recognition under domain shift}
\shorttitle{Approach for MIDOG 2025}

\author[1]{Gennaro Percannella}
\author[1]{Mattia Sarno}
\author[1]{Francesco Tortorella}
\author[1]{Mario Vento}
\affil[1]{Department of Information and Electrical Engineering and Applied Mathematics (DIEM), University of Salerno, Via Giovanni Paolo II
132, Fisciano, 84084, Salerno, Italy}

\maketitle

\begin{abstract}
Recognizing atypical mitotic figures in histopathology images allows physicians to correctly assess tumor aggressiveness. 
Although machine learning models could be exploited for automatically performing such a task, under domain shift these models suffer from significative performance 
drops. In this work, an approach based on multi-task learning is proposed for addressing this problem. By exploiting auxiliary tasks, correlated 
to the main classification task, the proposed approach, submitted to the track 2 of the \textbf{MI}tosis \textbf{DO}main \textbf{G}eneralization (MIDOG) challenge,
aims to aid the model to focus only on the object to classify, ignoring the domain varying background of 
the image. The proposed approach shows promising performance in a preliminary evaluation conducted on three distinct datasets, i.e., the MIDOG 2025 Atypical Training Set, the AMi-Br dataset, as well as the preliminary test
set of the MIDOG25 challenge.
\end{abstract}
\begin{corrauthor}
masarno@unisa.it
\end{corrauthor}

\section*{Introduction}
Characterizing mitotic figures in histopathology images in order to detect atypical mitoses, i.e., cells that do not follow
a structured division pattern, allows physicians to have an indicator of the aggressiveness of a cancer. Deep 
neural networks can be exploited for aiding physicians in such a classification task. However, as these models are generally trained under
the assumption that training and test data are sampled from the same domain, they fail in generalizing on data from domains that are different
from the training ones \cite{jahanifar2025domain}. In the context of histopathology, domain shift across datasets may occur due to several factors, such as variations 
in the image acquisition scanner or laboratory, in the analysed tissue or tumor type, as well as in the species which subjects belong to.\\
The \textbf{MI}tosis \textbf{DO}main \textbf{G}eneralization (MIDOG) challenge \cite{ammeling_mitosis_2025} addresses the problem of developing machine learning models robust to domain shift for mitosis detection and characterization in histopathology images. 
In particular, the track 2 of this challenge deals with the problem of characterizing mitotic figures as typical or atypical mitoses. 
To address this problem, in this work, we propose an approach that exploits multi-task learning (MTL) as a strategy for increasing model robustness against domain shift.

\section*{Materials and methods}
The proposed approach consists in a multi-task neural network that exploits, at training time, auxiliary tasks, as a regularization strategy against domain shift. 
Specifically, two auxilary dense-classification tasks correlated to the main task, i.e., mitosis binary segmentation and pixel-level mitosis classification, are exploited in the proposed MTL approach: the motivation is that, 
as such dense classification tasks explicitely provide the model with information about the localization of the object to classify, they could make an implict regularising contribution to the training process, 
aiding the model to focus only on the object of interest, regardless of possible variations in the image background, caused for instance by variations in the analysed tissue type across domains. In the following subsections, details about
the proposed method are provided.

\subsection*{Mitosis segmentation mask extraction}
In order to provide to our multi-task neural network supervision signals for carrying out the auxiliary dense-classification tasks, a ground truth segmentation mask for each mitosis has been extracted, 
by means of the segmentation model NuClick \cite{ALEMIKOOHBANANI2020101771}. This is an open-source interactive segmentation model, capable of segmenting cells in histopathology images, 
based on point annotations related to the cells to be segmented.\\ In this work, the coordinates of the center of each mitosis within the original \textit{.tiff} files have been extracted from the metadata of the provided datasets,
and they have been passed to NuClick for extracting the mitosis segmentation mask in such images.

\begin{figure*}[t]
  \centering
  \includegraphics[width=0.95\textwidth]{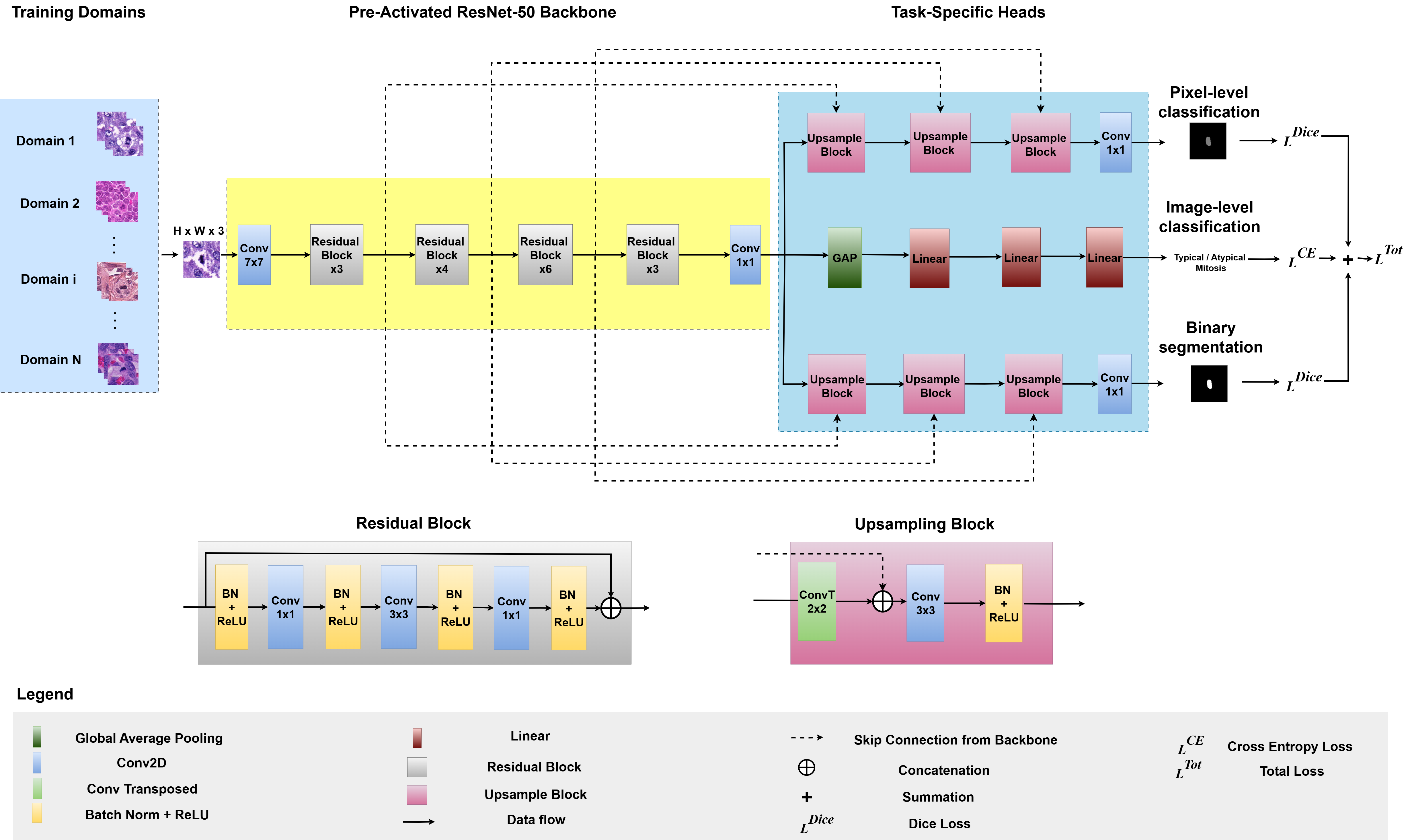}
  \caption{The proposed multi-task network}
  \label{fig:wide_image}
\end{figure*}

\subsection*{Network architecture}
In Figure \ref{fig:wide_image}, the proposed multi-task neural network is shown. It consists of a backbone, shared across the performed tasks, followed by three task-specific heads. As the network backbone, the Pre-Activated ResNet-50 presented in \cite{GRAHAM2019101563} has been exploited,
due to its capability of preserving, during feature extraction, the information needed for nuclei localization within histopathology images. Following the Pre-Activated ResNet-50 backbone, the proposed network consists of three task-specific heads, 
for carrying out the three addressed tasks, i.e., the main image-level mitosis classification task, and the auxiliary mitosis binary segmentation and pixel-level mitosis classification.
As regards the network head for the main task, it consists of three stacked fully connected layers, interleaved by ReLU activation functions.
Concerning instead the task-specific heads for the two auxiliary tasks, they have been designed as U-Net-like decoders: each decoder, in particular, consists of three up-sampling blocks, each of which upsamples the spatial dimension of the
input feature maps by a factor of 2. Skip connections are further exploited for allowing each decoder to better recover spatial details in the upsampling operations.

\subsection*{Training procedure}
The proposed network was trained for 50 epochs, using Adam as optimizer with learning rate $4\times10^{-5}$ and batch size equal to 24. As a loss function, the sum of three losses, one for each addressed task has been exploited: specifically,
a weighted binary cross entropy, with weights inversely proportional to class frequency, was exploited for carrying out the main classification task, while Dice loss \cite{sudre2017generalised} was used for both the auxiliary tasks.\\
To increase the variability of the training dataset, online data augmentation has been exploited, through random rotations, random flips, as well as stain augmentation.

\subsection*{Inference}
At inference time, task-specific heads associated to auxiliary tasks are pruned from the network, so as to allow it to perform only the main classification task.\\
To further improve model generalization capability, model ensembling and test-time-augmentation have been included in the inference pipeline: on one side, the predictions of the best found models in our
leave-one-domain out cross validation scheme are combined according to a major voting aggregation scheme; on the other hand, the predictions of each model on different perturbed view of the analysed images are averaged to improve its robustness.

\section*{Results}

\subsection*{Dataset}
The proposed approach was evaluated on two datasets, i.e., MIDOG 2025 Atypical Training Set and AMi-Br. The MIDOG 2025 Atypical Training Set \cite{aubreville2023comprehensive} consists of atypical mitotic figure subclassification for the entire MIDOG++ dataset, encompassing 11,939 mitotic figures from all 7 domains of MIDOG++.
Each domain, in particular, consists of images acqured through a certain scanner in a certain center, and referring to a certain tumor type in a a subject belonging to a certain species.
\\
The AMi-Br \cite{bertram2025histologic} dataset containts atypical mitotic figure subclassification for the MIDOG 2021 and the TUPAC16 challenge datasets \cite{aubreville2023mitosis}\cite{veta2015assessment}. 
The dataset consists of 3,720 annotated mitoses, split into 832 atypical and 2888 normal mitotic figures. In the conducted experiments, this dataset has been exploited as an additional test set, to further evaluate the robustness of the proposed method. In particular, 
only the portion of this dataset belonging to the TUPAC dataset has been exploited, in order to avoid superpositions with the images characterizing the MIDOG 2025 Atypical Training Set.

\subsection*{Experimental Protocol}
Experiments have been performed according to a leave-one-domain-out protocol, carried out on the MIDOG 2025 Atypical Training Set. Specifically, in each experiment, 
the domains within this dataset have been splitted into 5 training domains, 1 validation domain, and 1 test domain. Model selection was performed based on the value of loss on the validation domain.\\ 
In the conducted experiments, model performance has been assessed by measuring model balanced accuracy over each of the test sets of the leave-one-domain-out protocol, as well as on the AMi-Br dataset.

\begin{table}[h!]
\centering
\footnotesize
\begin{tabular}{|c|c|c|c|}  
\hline
\textbf{Model} & \textbf{Validation} & \textbf{Test} & \textbf{Test AMi-Br} \\
\hline
\textbf{Single task} & $0.842_{\pm 0.049}$ & $0.819_{\pm 0.067}$ & $0.793_{\pm 0.013}$ \\
\hline
\textbf{MTL} & $\textbf{0.847}_{\pm 0.046}$ & $\textbf{0.833}_{\pm 0.051}$ & $\textbf{0.806}_{\pm 0.013}$ \\
\hline
\end{tabular}
\caption{Mean and standard deviation of the balanced accuracy evaluated on the validation and test domains, as well as on the dataset AMi-Br, across the leave-one-domain-out experiments}
\label{tab:table exp}
\end{table}

\subsection*{Experimental Results}
The performance of the proposed method on both validation and test sets of leave-one-domain-out experiments, as well as on the AMi-Br test set are reported in Table \ref{tab:table exp}. As can be observed, the use of a MTL strategy for dealing with domain shift positively
contributes to model performance on data belonging to different domains from the training ones.\\
On the preliminary test set of MIDOG25 track 2, instead, the proposed approach achieves a balanced accuracy of 0.875, confirming its promising generalization capability under domain shift. 

\section*{Discussion}
In this work, an MTL approach has been proposed for addressing the problem of atypical mitosis recognition in histopathology images, under domain shift. The proposed approach, in particular, exploits at training time auxiliary dense-classification
tasks for reducing model sensitivity to variations in the image background, caused for instance by variations in the analysed tissue or tumor type across domains.\\ In the conducted experiments on two distinct datasets, 
as well as in the evaluation conducted on the preliminary test set of the track 2 of the MIDOG25 challenge, the proposed approach shows promising performance.

\section*{Bibliography}

\bibliography{literature.bib}

\begin{thebibliography}{9}
\providecommand{\natexlab}[1]{#1}
\providecommand{\url}[1]{\texttt{#1}}
\expandafter\ifx\csname urlstyle\endcsname\relax
  \providecommand{\doi}[1]{doi: #1}\else
  \providecommand{\doi}{doi: \begingroup \urlstyle{rm}\Url}\fi

\bibitem[Jahanifar et~al.(2025)Jahanifar, Raza, Xu, Vuong, Jewsbury, Shephard,
  Zamanitajeddin, Kwak, Raza, Minhas, et~al.]{jahanifar2025domain}
Mostafa Jahanifar, Manahil Raza, Kesi Xu, Trinh Thi~Le Vuong, Robert Jewsbury,
  Adam Shephard, Neda Zamanitajeddin, Jin~Tae Kwak, Shan E~Ahmed Raza, Fayyaz
  Minhas, et~al.
\newblock Domain generalization in computational pathology: survey and
  guidelines.
\newblock \emph{ACM Computing Surveys}, 57\penalty0 (11):\penalty0 1--37, 2025.

\bibitem[Ammeling et~al.(2025)Ammeling, Aubreville, Banerjee, Bertram,
  Breininger, Hirling, Horvath, Stathonikos, and Veta]{ammeling_mitosis_2025}
Jonas Ammeling, Marc Aubreville, Sweta Banerjee, Christof~A. Bertram, Katharina
  Breininger, Dominik Hirling, Peter Horvath, Nikolas Stathonikos, and Mitko
  Veta.
\newblock Mitosis {Domain} {Generalization} {Challenge} 2025.
\newblock Zenodo, March 2025.
\newblock \doi{10.5281/zenodo.15077361}.

\bibitem[{Alemi Koohbanani} et~al.(2020){Alemi Koohbanani}, Jahanifar, {Zamani
  Tajadin}, and Rajpoot]{ALEMIKOOHBANANI2020101771}
Navid {Alemi Koohbanani}, Mostafa Jahanifar, Neda {Zamani Tajadin}, and Nasir
  Rajpoot.
\newblock Nuclick: A deep learning framework for interactive segmentation of
  microscopic images.
\newblock \emph{Medical Image Analysis}, 65:\penalty0 101771, 2020.
\newblock ISSN 1361-8415.
\newblock \doi{https://doi.org/10.1016/j.media.2020.101771}.

\bibitem[Graham et~al.(2019)Graham, Vu, Raza, Azam, Tsang, Kwak, and
  Rajpoot]{GRAHAM2019101563}
Simon Graham, Quoc~Dang Vu, Shan E~Ahmed Raza, Ayesha Azam, Yee~Wah Tsang,
  Jin~Tae Kwak, and Nasir Rajpoot.
\newblock Hover-net: Simultaneous segmentation and classification of nuclei in
  multi-tissue histology images.
\newblock \emph{Medical Image Analysis}, 58:\penalty0 101563, 2019.
\newblock ISSN 1361-8415.
\newblock \doi{https://doi.org/10.1016/j.media.2019.101563}.

\bibitem[Sudre et~al.(2017)Sudre, Li, Vercauteren, Ourselin, and
  Jorge~Cardoso]{sudre2017generalised}
Carole~H Sudre, Wenqi Li, Tom Vercauteren, Sebastien Ourselin, and
  M~Jorge~Cardoso.
\newblock Generalised dice overlap as a deep learning loss function for highly
  unbalanced segmentations.
\newblock In \emph{International Workshop on Deep Learning in Medical Image
  Analysis}, pages 240--248. Springer, 2017.

\bibitem[Aubreville et~al.(2023{\natexlab{a}})Aubreville, Wilm, Stathonikos,
  Breininger, Donovan, Jabari, Veta, Ganz, Ammeling, van Diest,
  et~al.]{aubreville2023comprehensive}
Marc Aubreville, Frauke Wilm, Nikolas Stathonikos, Katharina Breininger,
  Taryn~A Donovan, Samir Jabari, Mitko Veta, Jonathan Ganz, Jonas Ammeling,
  Paul~J van Diest, et~al.
\newblock A comprehensive multi-domain dataset for mitotic figure detection.
\newblock \emph{Scientific data}, 10\penalty0 (1):\penalty0 484,
  2023{\natexlab{a}}.

\bibitem[Bertram et~al.(2025)Bertram, Weiss, Donovan, Banerjee, Conrad,
  Ammeling, Klopfleisch, Kaltenecker, and Aubreville]{bertram2025histologic}
Christof~A Bertram, Viktoria Weiss, Taryn~A Donovan, Sweta Banerjee, Thomas
  Conrad, Jonas Ammeling, Robert Klopfleisch, Christopher Kaltenecker, and Marc
  Aubreville.
\newblock Histologic dataset of normal and atypical mitotic figures on human
  breast cancer (ami-br).
\newblock In \emph{BVM Workshop}, pages 113--118. Springer, 2025.

\bibitem[Aubreville et~al.(2023{\natexlab{b}})Aubreville, Stathonikos, Bertram,
  Klopfleisch, Ter~Hoeve, Ciompi, Wilm, Marzahl, Donovan, Maier,
  et~al.]{aubreville2023mitosis}
Marc Aubreville, Nikolas Stathonikos, Christof~A Bertram, Robert Klopfleisch,
  Natalie Ter~Hoeve, Francesco Ciompi, Frauke Wilm, Christian Marzahl, Taryn~A
  Donovan, Andreas Maier, et~al.
\newblock Mitosis domain generalization in histopathology images—the midog
  challenge.
\newblock \emph{Medical Image Analysis}, 84:\penalty0 102699,
  2023{\natexlab{b}}.

\bibitem[Veta et~al.(2015)Veta, Van~Diest, Willems, Wang, Madabhushi, Cruz-Roa,
  Gonzalez, Larsen, Vestergaard, Dahl, et~al.]{veta2015assessment}
Mitko Veta, Paul~J Van~Diest, Stefan~M Willems, Haibo Wang, Anant Madabhushi,
  Angel Cruz-Roa, Fabio Gonzalez, Anders~BL Larsen, Jacob~S Vestergaard,
  Anders~B Dahl, et~al.
\newblock Assessment of algorithms for mitosis detection in breast cancer
  histopathology images.
\newblock \emph{Medical image analysis}, 20\penalty0 (1):\penalty0 237--248,
  2015.

\end{thebibliography}

\end{document}